\documentclass[onecolumn,preprintnumbers]{revtex4}
\usepackage{amsmath,amssymb,graphics,epsfig,subfigure}
\usepackage{color}
\usepackage[colorlinks,linkcolor=red,anchorcolor=red,citecolor=green]{hyperref}
\usepackage{setspace}
\usepackage{booktabs}
\usepackage{multirow}
\usepackage{diagbox}
\begin{document}

\thispagestyle{empty}

\begin{center}

\title{Thermodynamics curvature in phase transitions for AdS black hole}

\author{Zhen-Ming Xu\footnote{xuzhenm@nwu.edu.cn}, Bin Wu, and Wen-Li Yang
        \vspace{6pt}\\}

\affiliation{ $^{1}$Institute of Modern Physics, Northwest University, Xi'an 710127, China\\
$^{2}$School of Physics, Northwest University, Xi'an 710127, China\\
$^{3}$Shaanxi Key Laboratory for Theoretical Physics Frontiers, Xi'an 710127, China\\
$^{4}$Peng Huanwu Center for Fundamental Theory, Xi'an 710127, China}

\begin{abstract}
We investigate the thermodynamic curvature in the Hawking-Page phase transition and the second-order phase transition of the AdS black hole. It is shown that the thermodynamic curvature has the same behavior in these two different phase transitions. Specifically, the thermodynamic curvature in the Hawking-Page phase transition is the power function of the Hawking-Page phase transition temperature with exponent related to the spacetime dimension, or is invariably proportional to the reciprocal of the Hawking-Page phase transition entropy. For the thermodynamic curvature in the second-order phase transition, the conclusion is similar. The more enlightening result is that the ratio of thermodynamic curvatures along the two different phase transitions tends to the natural constant in the limit that the number of spacetime dimensions is infinite. The universal and novel result provides an essential difference between the Hawking-Page phase transition and the second-order phase transition of the AdS black hole in the large dimension paradigm.
\end{abstract}

\maketitle
\end{center}

\section{Introduction}
Phase transition has always been an important issue in theoretical physics. Especially when the black hole is mapped to the thermodynamic system, the analysis of the phase transition of the black hole can help us to explore the connotation of quantum gravity. Among them, some common characteristics and essential differences of different phase transitions of black holes are particularly important. This universal feature will bring us the most profound understanding of a physical system, just like the critical exponent and critical amplitude in the theory of phase transition, where the universal constant properties satisfied by different systems have revolutionized our understanding of the phase transition theory. In our current research, we find that there are some common characteristics between the two different phase transitions of the AdS black hole system from the perspective of thermodynamic geometry. At the same time, more importantly, the essential differences between the two different phase transitions can be reflected by the most important constant in nature, the natural constant. This result may be very important for us to understand the phase transition of black hole and the physical connotation of thermodynamic geometry.

Thermodynamic geometry now is regarded as an effective exploration of mathematical language description of thermodynamics theory. Weinhold~\cite{Weinhold1975} introduced the thermodynamic geometry for the first time, where the internal energy is chosen to be the thermodynamic potential. Unfortunately this geometry seems physically meaningless in the context of purely equilibrium thermodynamics. Subsequently, based on the equilibrium fluctuation theory, Ruppeiner~\cite{Ruppeiner1995} further developed the geometric theory of thermodynamics, and put forward a thermodynamic metric with the entropy as the thermodynamic potential. The components of the inverse Ruppeiner thermodynamic metric give second moments of fluctuations. This makes the Ruppeiner thermodynamic geometry get extensive attention~\cite{Cai1999,Zhang2015a,Liu2010,Dimov2019,Wei2015,Miao2018a,Zangeneh2018,Xu2019,Xu2020a,Ghosh2020,Wei2019a,Wei2019b}.

For ordinary fluid systems, a large number of studies suggest that Ruppeiner thermodynamic curvature is related to the correlation length $\xi$ of the thermodynamic system via $R\sim \kappa  \xi^d$, where $\kappa$ is a dimensionless constant, and $d$ denotes the  physical dimensionality of the system~\cite{Ruppeiner2010}. This conclusion seems to be potentially indicated that the Ruppeiner geometry can be exploited to probe the microstructure of a thermodynamic system from its macroscopic thermodynamic quantities (it can be regarded as the inverse process of statistical physics).

For a black hole system, due to the discovery of the Hawking temperature and Beckenstein-Hawking entropy, it has entered the field of thermodynamics~\cite{Hawking1975,Bekenstein1973}. In particular, the introduction of extended phase space, that is, the negative cosmological constant is treated as thermodynamic pressure~\cite{Kastor2009,Dolan2011}, makes some novel thermodynamic behaviors of black holes gradually revealed. The most famous one is the Hawking-Page phase transition, which means that a stable large black hole can exchange energy and establish the equilibrium with the thermal anti-de Sitter (AdS) background~\cite{Hawking1983}. Subsequently Witten~\cite{Witten1998} established the holographic duality between the Hawking-Page phase transition in AdS black hole and the confinement/deconfinement phase transition in gauge theory.

In addition, the AdS black hole also has a very significant characteristic with a minimum temperature. At this minimum, the heat capacity at constant pressure diverges. Davies~\cite{Davies1977,Davies1978} argued that the divergence of the heat capacity corresponds to the second-order phase transition between unstable state and stable state. Hence the transition point is also called Davies point. Moreover, a complete analogy between the van der Waals fluid and the charged AdS black hole~\cite{Kubiznak2012}, black hole as a holographic heat engine~\cite{Johnson2014}, and some other interesting topics have been widely studied and discussed~\cite{Kubiznak2017}. Naturally, the idea of thermodynamic geometry is also used in black hole system~\cite{Ruppeiner2014}. On the one hand, it can be used to analyze some phase transition behavior of the black hole~\cite{Cai1999,Zhang2015a,Liu2010,Dimov2019}. On the other hand, based on the hypothesis of the black hole molecular~\cite{Wei2015}, it can also be used to explore the possible microstructure of the black hole~\cite{Miao2018a,Zangeneh2018,Xu2019,Xu2020a,Ghosh2020,Wei2019a,Wei2019b}. Furthermore, some attempts have also been made to explore the counterpart of dual conformal field theory in thermodynamic geometry~\cite{Zhang2015b,Bhamidipati2020}.

Because of the importance of phase transition in black hole thermodynamics, our current interest is to investigate some universal characteristics of thermodynamic curvature along the phase transition of the AdS black hole. On the one hand, the significance of the Hawking-Page phase transition is very clear both in the gravity and the field theory. In the gravity side, it indicates that a stable large black hole can exchange energy and establish the equilibrium with the thermal AdS background. In the field theory side, it corresponds to the confinement/deconfinement phase transition. For the second-order phase transition, it can reflect the thermal stability of the AdS black hole. These two kinds of phase transitions are completely different in nature. In addition, as mentioned earlier, the thermodynamics curvature is also an important physical quantity in the theory of black hole thermodynamics. Hence it is helpful to understand the meaning of thermodynamics geometry in black hole by studying the behavior of thermodynamic curvature in the Hawking-Page phase transition and the second-order phase transition. On the other hand, recent study~\cite{Wei2020} pointed out that the normalized thermodynamic curvature~\cite{Wei2019c} is a universal constant at the Hawking-Page transition point. It is a very meaningful result. In this paper, we study the behavior of the thermodynamic curvature in the two different phase transition to find some common characteristics and essential differences of these two phase transitions. Here we need to explain that the normalized thermodynamic curvature is completely different from the thermodynamic curvature in the current paper. For details, see Section \ref{sec2}. Moreover, the thermodynamic curvatures induced by different thermodynamic potentials are equivalent to each other~\cite{Xu2020a,Xu2020b}. In other words, thermodynamic curvature is independent of thermodynamic coordinates. It is also the natural result of the Legendre transformation between different thermodynamic potentials.

The results show that the thermodynamic curvature along the Hawking-Page phase transition curve is the power function of the Hawking-Page temperature with exponent $(n-2)$, where $n$ is the number of spacetime dimension. The same conclusion is obtained for the second-order phase transition. At the same time, we also find that the ratio of the thermodynamic curvature along the curves of the Hawking-Page phase transition and the second-order phase transition is a constant only related to dimension $n$. A universal and novel result is that the ratio tends to the natural constant $e$ when we consider the large dimension limit. It provides an essential difference between the Hawking-Page phase transition and the second-order phase transition of the AdS black hole in the large dimension paradigm. Meanwhile this result also reflects some unique characteristics of thermodynamic curvature in black hole physics. It is of great significance for us to further understand thermodynamics geometry in gravitational theory and dual field theory.

\section{Thermodynamic curvature}\label{sec2}
Based on the equilibrium fluctuation theory of thermodynamics, Ruppeiner further deduces the interesting content beyond the equilibrium fluctuation theory in the standard textbook, that is, the theory later called the Ruppeiner thermodynamic geometry. This is a new attempt to describe the thermodynamic theory in geometric language. Accordingly, the thermodynamic metric (or line element) is expressed as a Hessian structure with entropy as the thermodynamic potential~\cite{Ruppeiner1995}
\begin{equation}\label{line}
\Delta l^2=g_{\mu\nu}\Delta X^\mu \Delta X^\nu=-\frac{\partial^2 S}{\partial X^\mu \partial X^\nu}\Delta X^\mu \Delta X^\nu,
\end{equation}
where $X^{\mu}$ correspond to some extensive variables of the system. In the light of Eq.~(\ref{line}), by analogy with geometric curvature, one can naturally use the Christoffel symbols
$\Gamma^{\alpha}_{\beta\gamma}=\frac12g^{\mu\alpha}\left(\partial_{\gamma}g_{\mu\beta}+
\partial_{\beta}g_{\mu\gamma}-\partial_{\mu}g_{\beta\gamma}\right)$
and the Riemannian curvature tensor
${R^{\alpha}}_{\beta\gamma\delta}=\partial_{\delta}\Gamma^{\alpha}_{\beta\gamma}-\partial_{\gamma}\Gamma^{\alpha}_{\beta\delta}+
\Gamma^{\mu}_{\beta\gamma}\Gamma^{\alpha}_{\mu\delta}-\Gamma^{\mu}_{\beta\delta}\Gamma^{\alpha}_{\mu\gamma}$ to obtain the thermodynamic (scalar) curvature $R=g^{\mu\nu}{R^{\xi}}_{\mu\xi\nu}$~\cite{Ruppeiner1995}.

However, for the AdS black hole, we frequently know the expression of the mass of the black hole, i.e., $M=M(S,P,...)$. In this event, simplification results on writing the thermodynamic metric in the other thermodynamic potential are need. For this purpose, there are currently two different options. One is based on the differential relationship of internal energy~\cite{Wei2019c}, and the other is directly based on the mass (or formally analogous to enthalpy) of the AdS black hole~\cite{Xu2020a}.

\subsection{Internal energy: $dU=TdS-PdV$}
Here we know that the internal energy $U$, entropy $S$ and thermodynamic volume $V$ are extensive variables, while the temperature $T$ and pressure $P$ are intensive variables. Considering the thermodynamic metric~(\ref{line}), we need to write
\begin{eqnarray}
dS=\frac{1}{T}dU+\frac{P}{T}dV.
\end{eqnarray}
Now we can take $X^{\mu}=(U,V)$ and the intensive variables corresponding to $X^{\mu}$ are $Y_{\mu}=\partial S/\partial X^{\mu}=(1/T,P/T)$. After some simple calculations, the line element~(\ref{line}) becomes a universal form~\cite{Wei2019a,Wei2019b}
\begin{eqnarray}\label{gmetricu}
\Delta l^2_{_{U}}=-\Delta Y_{\mu} \Delta X^{\mu}=\frac{1}{T}\Delta T \Delta S-\frac{1}{T}\Delta V \Delta P.
\end{eqnarray}
According to the expression $dU=TdS-PdV$, we can clearly see that there are four coordinate spaces $\{S, P\}$, $\{T, V\}$, $\{S, V\}$ and $\{T, P\}$ to re-express the thermodynamic metric in the other thermodynamic potential.
\begin{itemize}
  \item In the coordinate space $\{S, P\}$, namely the enthalpy as the thermodynamic potential
    \begin{eqnarray}\label{ulinesp}
\Delta l^2_{_{U}} =\frac{1}{T}\left(\frac{\partial T}{\partial S}\right)_P \Delta S^2-\frac{1}{T}\left(\frac{\partial V}{\partial P}\right)_S \Delta P^2=g_{\mu\nu}\Delta x^{\mu}\Delta x^{\nu} \qquad (x^{\mu}=S,P).
\end{eqnarray}
  \item In the coordinate space $\{T, V\}$, namely the Helmholtz free energy as the thermodynamic potential
    \begin{eqnarray}\label{ulinetv}
\Delta l^2_{_{U}}=\frac{1}{T}\left(\frac{\partial S}{\partial T}\right)_V \Delta T^2-\frac{1}{T}\left(\frac{\partial P}{\partial V}\right)_T \Delta V^2=g_{\mu\nu}\Delta x^{\mu}\Delta x^{\nu} \qquad (x^{\mu}=T,V).
\end{eqnarray}
  \item In the coordinate space $\{S, V\}$, namely the internal energy as the thermodynamic potential
    \begin{eqnarray}\label{ulinesv}
   \begin{aligned}
    \Delta l^2_{_{U}} &=\frac{1}{T}\left(\frac{\partial T}{\partial S}\right)_V \Delta S^2+\frac{2}{T}\left(\frac{\partial T}{\partial V}\right)_S \Delta S \Delta V-\frac{1}{T}\left(\frac{\partial P}{\partial V}\right)_S \Delta V^2\\
    &=g_{\mu\nu}\Delta x^{\mu}\Delta x^{\nu} \qquad (x^{\mu}=S,V).
  \end{aligned}
  \end{eqnarray}
  \item In the coordinate space $\{T, P\}$, namely the Gibbs free energy as the thermodynamic potential
   \begin{eqnarray}\label{ulinetp}
\begin{aligned}
\Delta l^2_{_{U}} &=\frac{1}{T}\left(\frac{\partial S}{\partial T}\right)_P \Delta T^2+\frac{2}{T}\left(\frac{\partial S}{\partial P}\right)_T \Delta T \Delta P-\frac{1}{T}\left(\frac{\partial V}{\partial P}\right)_T \Delta P^2\\
&=g_{\mu\nu}\Delta x^{\mu}\Delta x^{\nu} \qquad (x^{\mu}=T,P).
\end{aligned}
\end{eqnarray}
\end{itemize}

In principle, the thermodynamic curvatures, $R^{(U)}$, obtained by the above four line elements are equivalent to each other due to the Legendre transformation between the different thermodynamic potential. However for the (charged) Schwarzschild AdS black hole, because the thermodynamic volume and entropy are not independent of each other, i.e., $S=S(V)$ or $V=V(S)$, it renders the above four line elements singular. For example, in the coordinate space $\{T, V\}$, the thermodynamic curvature behaviors as $R^{(U)}\varpropto 1/C_V$ where the heat capacity at constant volume $C_{_V}:=T(\partial S/\partial T)_{_V}$. Obviously $C_V=0$ causes thermodynamic curvature $R^{(U)}$ to diverge. To cure this singularity, authors in~\cite{Wei2019c} have introduced the normalized thermodynamic curvature $R_N = R^{(U)} C_V$ from the thermodynamic curvature $R^{(U)}$. In this way, we can extract some interesting thermodynamic properties of black holes from a finite normalized thermodynamic curvature.

\subsection{Mass: $dM=TdS+VdP$}
Here the mass $M$, entropy $S$ and thermodynamic pressure $P$ are extensive variables, while the temperature $T$ and thermodynamic volume $V$ are intensive variables. Considering the thermodynamic metric~(\ref{line}), we need to write
\begin{eqnarray}
dS=\frac{1}{T}dM-\frac{V}{T}dP.
\end{eqnarray}
Taking $X^{\mu}=(M,P)$, then we have the intensive variables corresponding to $X^{\mu}$ are $Y_{\mu}=\partial S/\partial X^{\mu}=(1/T,-V/T)$. Hence the line element~(\ref{line}) becomes a universal form~\cite{Xu2020a}
\begin{eqnarray}\label{gmetricm}
\Delta l^2_{_{M}}=-\Delta Y_{\mu} \Delta X^{\mu}=\frac{1}{T}\Delta T \Delta S+\frac{1}{T}\Delta V \Delta P.
\end{eqnarray}
According to the expression $dM=TdS+VdP$, we can clearly see that there are four coordinate spaces $\{S, P\}$, $\{T, V\}$, $\{S, V\}$ and $\{T, P\}$ to re-express the thermodynamic metric in the other thermodynamic potential.
\begin{itemize}
  \item In the coordinate space $\{S, P\}$, namely the enthalpy as the thermodynamic potential
  \begin{eqnarray}\label{mlinesp}
   \begin{aligned}
    \Delta l^2_{_{M}} &=\frac{1}{T}\left(\frac{\partial T}{\partial S}\right)_P \Delta S^2+\frac{2}{T}\left(\frac{\partial T}{\partial P}\right)_S \Delta S \Delta P+\frac{1}{T}\left(\frac{\partial V}{\partial P}\right)_S \Delta P^2\\
    &=g_{\mu\nu}\Delta x^{\mu}\Delta x^{\nu} \qquad (x^{\mu}=S,P).
  \end{aligned}
  \end{eqnarray}
  \item In the coordinate space $\{T, V\}$, namely the Helmholtz free energy as the thermodynamic potential
  \begin{eqnarray}\label{mlinetv}
\begin{aligned}
\Delta l^2_{_{M}} &=\frac{1}{T}\left(\frac{\partial S}{\partial T}\right)_V \Delta T^2+\frac{2}{T}\left(\frac{\partial S}{\partial V}\right)_T \Delta T \Delta V+\frac{1}{T}\left(\frac{\partial P}{\partial V}\right)_T \Delta V^2\\
&=g_{\mu\nu}\Delta x^{\mu}\Delta x^{\nu} \qquad (x^{\mu}=T,V).
\end{aligned}
\end{eqnarray}
  \item In the coordinate space $\{S, V\}$, namely the internal energy as the thermodynamic potential
  \begin{eqnarray}\label{mlinesv}
\Delta l^2_{_{M}} =\frac{1}{T}\left(\frac{\partial T}{\partial S}\right)_V \Delta S^2+\frac{1}{T}\left(\frac{\partial P}{\partial V}\right)_S \Delta V^2=g_{\mu\nu}\Delta x^{\mu}\Delta x^{\nu} \qquad (x^{\mu}=S,V).
\end{eqnarray}
  \item In the coordinate space $\{T, P\}$, namely the Gibbs free energy as the thermodynamic potential
  \begin{eqnarray}\label{mlinetp}
\Delta l^2_{_{M}}=\frac{1}{T}\left(\frac{\partial S}{\partial T}\right)_P \Delta T^2+\frac{1}{T}\left(\frac{\partial V}{\partial P}\right)_T \Delta P^2=g_{\mu\nu}\Delta x^{\mu}\Delta x^{\nu} \qquad (x^{\mu}=T,P).
\end{eqnarray}
\end{itemize}

The thermodynamic curvatures, $R^{(M)}$, obtained by the above four line elements are equivalent to each other due to the Legendre transformation between the different thermodynamic potential. For the black hole thermodynamic system, the above four line elements are always effective. In other words, there is no singularity although the thermodynamic volume and entropy are not independent of each other for the (charged) Schwarzschild AdS black hole. In this way, according to any one of the above four line elements, we can obtain a finite thermodynamic curvature $R^{(M)}$. For simplicity, the general form of thermodynamic curvature of the Schwarzschild AdS black hole in the coordinate space $\{S, P\}$ is~\cite{Miao2018b}
\begin{eqnarray}\label{gcur}
R^{(M)}=\frac{\partial}{\partial S}\left[\ln\left(T/\frac{\partial T}{\partial P}\right)\right].
\end{eqnarray}

\subsection{Some comments}
For two different options, we obtain two different thermodynamic curvatures $R^{(U)}$ or $R_N$ for the differential relationship of internal energy $dU=TdS-PdV$ as a basic starting point and $R^{(M)}$ for the differential relationship of the mass (or formally analogous to enthalpy) of the AdS black hole $dM=TdS+VdP$ as a basic starting point. Physically, the reason for this difference lies largely  in which of the pressure and volume we treat as the extensive variable. According to IUPAC~\cite{IUPAC}, an intensive quantity is one whose magnitude is independent of the size of the system, whereas an extensive quantity is one whose magnitude is additive for subsystems. Pressure is widely considered an intensive property and volume is extensive. Almost all textbooks are based on this definition. However, in some special situations, there may be a problem of duality between the intensive and extensive quantities.

In the thermodynamic statistical physics, we investigate the case that two ideal gases of the same pressure and volume merge into a new system. Which of the pressure and volume is extensive depends on the ways of merging of the two subsystems. If the pressure is kept constant during the merging, the volume of the total system is extensive and additive. Conversely if the volume is kept unchanged during the merging, the pressure of the total system is then extensive and additive.

Another very similar example is the voltage of the battery and the current in the circuit. If two batteries are connected in series, then the voltage multiplies and it is like the extensive variable; if two batteries are connected in parallel, then the current multiplies and it is like the extensive variable~\cite{Xu2020b}.

Moreover, in the textbook~\cite{Horsley1993}, the author write ``If I have some gas in a rigid container, I let half the gas out of the cylinder, then the pressure inside the cylinder will certainly fall and the pressure of the released gas will also be quite different from its initial value. Thus the pressure is an extensive property.''

Back to our current topic, two subsystems, the AdS black hole and its surroundings, make up an isolated system. This is very similar to the example of two ideal gases merging. Hence we can see that the thermodynamic curvature $R^{(U)}$ or $R_N$ is for the case that the thermodynamic pressure of a black hole remains unchanged when it forms an isolated system with its surroundings. While the thermodynamic curvature $R^{(M)}$ is for the case that the volume of a black hole remains unchanged when it forms an isolated system with its surroundings. In other words, these two different thermodynamic curvatures represent two different isolated systems in equilibrium. In~\cite{Wei2020}, authors obtain an interesting result that $R_N$ is a universal constant at the Hawking-Page transition point. In our present work, we can show later that $R^{(M)}$ along the Hawking-Page phase transition and the second-order phase transition is proportional to the reciprocal of the corresponding entropy and the ratio of $R^{(M)}$ along the two different phase transitions tends to the natural constant in the limit that the number of spacetime dimensions is infinite. In the following discussion, we ignore the superscript in $R^{(M)}$, abbreviated as $R:=R^{(M)}$.

\section{Thermodynamic curvature in the phase transition}\label{sec3}
Now we consider the $n$-dimensional Schwarzschild-Tangherlini AdS ($n$-STAdS) black hole. The metric is~\cite{Tangherlini1936,Belhaj2015}
\begin{equation}
d s^2=-f(r)dt^2+\frac{d r^2}{f(r)}+r^2 d\Omega_{n-2}^2,
\end{equation}
and the function $f(r)$ is
\begin{equation*}
f(r)=1-\frac{16\pi M}{(n-2)\omega r^{n-3}}+\frac{r^2}{l^2},
\end{equation*}
where $d\Omega_{n-2}^2$ is the square of line element on an $(n-2)$-dimensional unit sphere, $l$ represents the curvature radius of the AdS spacetime, and $M$ is the ADM mass of the black hole. In addition, the parameter $\omega$ denotes the area of an $(n-1)$-dimensional unit sphere and its value is related to the gamma function $\Gamma(x)$ via
$\omega=2\pi^{\frac{n-1}{2}}/\Gamma\left(\frac{n-1}{2}\right)$.

Naturally some importantly thermodynamic quantities of the $n$-STAdS black hole can be obtained in terms of the horizon radius $r_h$ which is the largest root of equation $f(r)=0$. The Hawking temperature and entropy of the black hole are~\cite{Belhaj2015}
\begin{eqnarray}\label{tands}
T=\frac{n-3}{4\pi r_h}+\frac{(n-1)r_h}{4\pi l^2}, \qquad S=\frac{\omega r_h^{n-2}}{4}.
\end{eqnarray}
The thermodynamic pressure and thermodynamic volume take the form~\cite{Belhaj2015}
\begin{eqnarray}\label{pandv}
P=\frac{(n-2)(n-1)}{16\pi l^2}, \qquad V=\frac{\omega r_h^{n-1}}{n-1}.
\end{eqnarray}
Due to the appearance of the above thermodynamic pressure and volume terms, the mass $M$ of the black hole corresponds to the enthalpy in thermodynamics instead of the internal energy, i.e., $H=M$. Furthermore, the more important physical quantity for analyzing the thermodynamic phase transition of the black hole is free energy. Specifically, for AdS black holes, it is Gibbs free energy, which is the Legendre transformation of the enthalpy~\cite{Belhaj2015},
\begin{eqnarray}\label{gibbs}
G\equiv H-TS=\frac{\omega}{16\pi}\left(r_h^{n-3}-\frac{r_h^{n-1}}{l^2}\right).
\end{eqnarray}

By inserting Eqs.~(\ref{tands}) and~(\ref{pandv}) into~(\ref{gcur}), we can obtain the thermodynamic curvature of $n$-STAdS black hole
\begin{eqnarray}\label{curbh}
R=-\frac{8(n-3)l^2 r_h^{2-n}}{(n-2)\omega[(n-3)l^2+(n-1)r_h^2]}.
\end{eqnarray}

\begin{figure}
\begin{center}
\includegraphics[width=80mm]{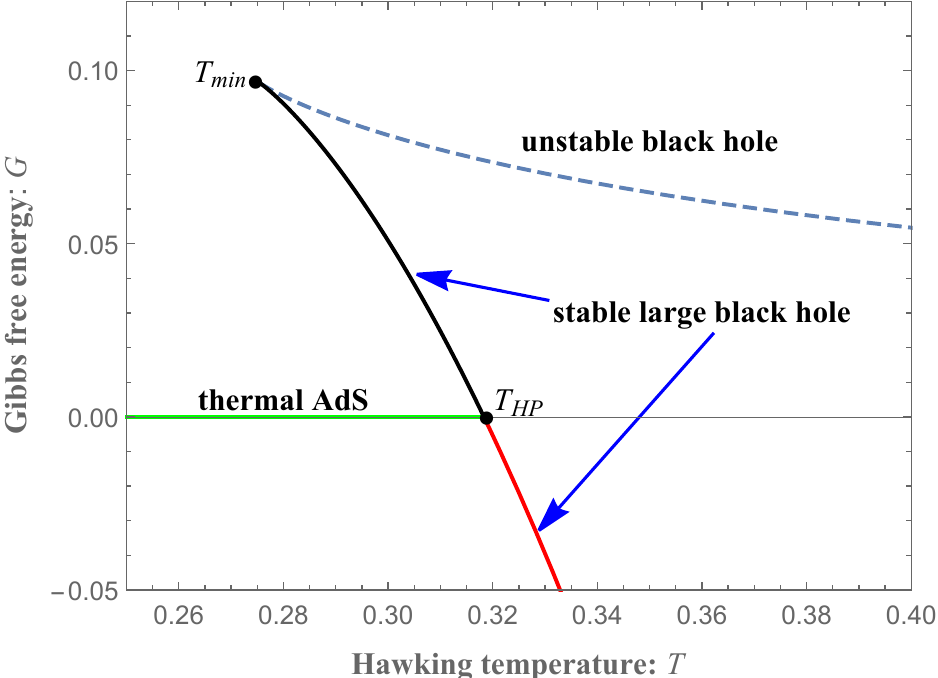}
\end{center}
\caption{The specific schematic diagram of the Hawking-Page phase transition for 4-dimensional Schwarzschild AdS black hole.}
\label{fig1}
\end{figure}

Because the thermal AdS background or the AdS black hole with the global minimum of Gibbs free energy is thermodynamically preferred respectively. A stable large black hole can exchange energy and establish the equilibrium with the thermal AdS background in the process of Hawking radiation from black holes, where the Gibbs free energy of the thermal AdS background is zero. The specific schematic diagram is shown in Fig.~\ref{fig1}. Hence the Hawking-Page phase transition satisfies the condition $G=0$, which means that $r_h^2=l^2$. In terms of Eqs.~(\ref{tands}) and~(\ref{pandv}), we finally obtain the entropy of the Hawking-Page phase transition
\begin{eqnarray}\label{hpts}
S_{\text{HP}}=4^{1-n}(n^2-3n+2)^{n/2-1}\pi^{1-n/2}\omega P^{1-n/2}.
\end{eqnarray}
Correspondingly, the Hawking-Page temperature is~\cite{Wei2020}
\begin{eqnarray}\label{hpt}
T_{\text{HP}}=\sqrt{\frac{4(n-2)P}{\pi(n-1)}}.
\end{eqnarray}
In particular, as $P$ has no terminal point in Eqs.~(\ref{hpts}) and~(\ref{hpt}), the Hawking-Page phase transition can happen at all pressures.

\begin{figure}
\begin{center}
\includegraphics[width=80mm]{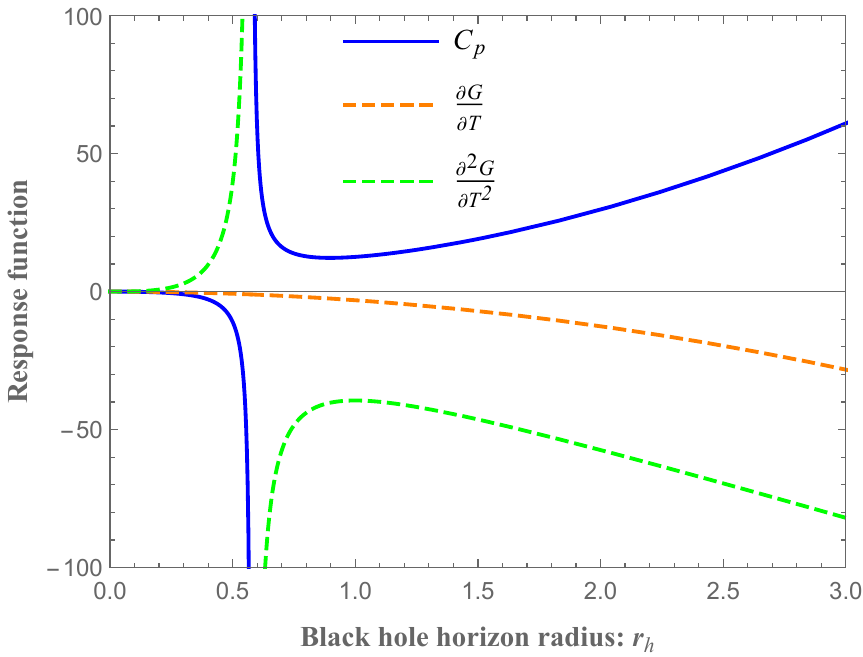}
\end{center}
\caption{The diagram of the second-order phase transition for 4-dimensional Schwarzschild AdS black hole. At $T=T_{\text{min}}$, i.e., $r_h=l/\sqrt{3}$, we can clearly see that the heat capacity at constant pressure is divergent, and the second derivative of the Gibbs free energy with respect to temperature diverge, while its first derivative to temperature is continuous.}
\label{fig2}
\end{figure}

Meanwhile, we also obtain the thermodynamic curvature of $n$-STAdS black hole along the Hawking-Page curve
\begin{eqnarray}\label{hpr}
R_{\text{HP}}=-\frac{n-3}{(n-2)^2 S_{\text{HP}}}=-\frac{(n-3)2^n \pi^{n-2}}{\omega(n-2)^n}T_{\text{HP}}^{n-2},
\end{eqnarray}
stating that the thermodynamic curvature along the curve of the Hawking-Page phase transition is proportional to the
reciprocal of the Hawking-Page phase transition entropy, or is the power function of the Hawking-Page temperature with exponent $(n-2)$.

Next we turn to another interesting property of the $n$-STAdS black hole. There is a minimum in the temperature curve, which satisfies the equation $\partial T/\partial r_h=0$, i.e., $(n-1)r_h^2=(n-3)l^2$. At this minimum, the heat capacity at constant pressure $C_P=(T \partial S/\partial T)_P$ of the black hole is divergent, which means that the black hole will experience the transition between the unstable phase and the stable phase in Fig.~\ref{fig2}. According to Ehrenfest's classification of phase transitions, we can observe that for the $n$-STAdS AdS black hole, when the temperature is taken as the thermodynamic order parameter, the Gibbs free energy itself and its first derivative to temperature are continuous, but the second derivative is discontinuous (see Figs.~\ref{fig1} and ~\ref{fig2}). Namely
\begin{eqnarray}\label{stp}
G\big|_{T=T_{\text{min}}} &=& \frac{2\omega}{16\pi(n-3)}\left(\frac{n-3}{n-1}\right)^{\frac{n-1}{2}}l^{n-3},\nonumber\\
\frac{\partial G}{\partial T}\bigg|_{T=T_{\text{min}}}&=& -\frac{\omega}{4}\left(\frac{n-3}{n-1}\right)^{\frac{n-2}{2}}l^{n-2},\\
\frac{\partial^2 G}{\partial T^2}\bigg|_{T=T_{\text{min}}}&\rightarrow & \pm \infty. \nonumber
\end{eqnarray}
Hence it is the second-order phase transition. In terms of Eqs.~(\ref{tands}) and~(\ref{pandv}), we finally obtain the entropy of the second-order phase transition,
\begin{eqnarray}\label{mts}
S_{T_{\text{min}}}=4^{1-n}(n^2-5n+6)^{n/2-1}\pi^{1-n/2}\omega P^{1-n/2}.
\end{eqnarray}
Note that the black hole is stable for $S>S_{T_{\text{min}}}$, and is unstable for $S<S_{T_{\text{min}}}$.
Correspondingly, the second-order phase transition temperature or the minimum temperature is~\cite{Wei2020}
\begin{eqnarray}\label{mt}
T_{\text{min}}=\sqrt{\frac{4(n-3)P}{\pi(n-2)}}.
\end{eqnarray}
As $P$ has no terminal point in Eqs.~(\ref{mts}) and~(\ref{mt}), the second-order phase transition always exists at all pressures. Moreover a very novel result that $T_{\text{HP}}|_{n=d}=T_{\text{min}}|_{n=d+1}$, i.e., the $d$-dimensional Hawking-Page temperature exactly equals to the black hole minimum temperature in one larger dimension, has been reported in Ref.~\cite{Wei2020}.

In addition, we also obtain the thermodynamic curvature of $n$-STAdS black hole along the curve of the second-order phase transition
\begin{eqnarray}\label{mr}
R_{T_{\text{min}}}=-\frac{1}{(n-2)S_{T_{\text{min}}}}=-\frac{2^n \pi^{n-2}}{\omega(n-2)(n-3)^{n-2}}T_{\text{min}}^{n-2}.
\end{eqnarray}
It is also proportional to the reciprocal of the black hole entropy of the second-order phase transition, or is the power function of the black hole minimum temperature with exponent $(n-2)$.

Interestingly, looking at Eqs.~(\ref{hpr}) and~(\ref{mr}), we find that the thermodynamic curvature has similar behavior under the Hawking-Page phase transition and the second-order phase transition, which allows us to get a novel result
\begin{eqnarray}\label{ratio}
\frac{R_{\text{HP}}}{R_{T_{\text{min}}}}=\frac{n-3}{n-2}\left(\frac{n-3}{n-1}\right)^{n/2-1},
\end{eqnarray}
namely that the ratio is a universal constant, independent of all thermodynamic parameters except the dimension. What's even more surprising is that if we take the large dimension limit, we have
\begin{eqnarray}\label{nratio}
\lim_{n\rightarrow\infty}\frac{R_{\text{HP}}}{R_{T_{\text{min}}}}=\frac{1}{e}.
\end{eqnarray}

Because the Hawking-Page phase transition and the second-order phase transition are very important properties of the AdS black hole. The ratio of thermodynamic curvature along the curves of the Hawking-Page phase transition and the second-order phase transition tends to the natural constant $e$ at $n\rightarrow\infty$. This unique constant connects two different phase transitions at the level of the thermodynamic geometry, which may reflect some universal characteristics of thermodynamic curvature in black hole physics.

\section{Discussion}\label{sec4}
We have shown that the thermodynamic curvature along the curves of the Hawking-Page phase transition and the second-order phase transition is always the power function of the corresponding temperature with exponent $(n-2)$, or equivalently, the thermodynamic curvature along the curves of two phase transitions is proportional to the reciprocal of the corresponding entropy respectively. This result shows that although the two phase transitions are different in nature, they have some common characteristics in the framework of the thermodynamic geometry. It is worth mentioning that the similar phenomenon about the thermodynamic curvature also appears in the case that some thermodynamic variables take the limit value. For example, the thermodynamic curvature of the charged black hole has a remanent of the reciprocal of entropy when the charge is zero~\cite{Mirza2007}, and the thermodynamic curvature of the extreme black hole of the super-entropic black hole has a remnant proportional to the reciprocal of entropy of the black hole~\cite{Xu2020c}. The comparison between our present results and those in Ref.~\cite{Wei2020} is shown in Table~\ref{tab}. The two different results fully reflect that the system we studied is completely different from the one considered in Ref.~\cite{Wei2020}.
\begin{table}[!h]
\centering
\begin{tabular}{c|c|c|c}
\hline
\multicolumn{2}{c|}{} &$R_N=R^{(U)} C_V$  &$R^{(M)}$ \\
\hline
\multicolumn{2}{c|}{{\diagbox[width=7cm,outerrightsep=25pt]{Item}{Value}{Expression}}}& $-\frac{n-3}{2}\frac{4\pi T r_h-n+3}{(2\pi T r_h-n+3)^2}$~\cite{Wei2020}  & $-\frac{8(n-3)l^2 r_h^{2-n}}{(n-2)\omega[(n-3)l^2+(n-1)r_h^2]}$\\
\hline
\rule{0pt}{18pt} At second-order phase transition: & \multirow{2}{*}{$R_{T_{\text{min}}}=$} &\multirow{2}{*}{$-\infty$}   &\multirow{2}{*}{$-\frac{2^n \pi^{n-2}}{\omega(n-2)(n-3)^{n-2}}T_{\text{min}}^{n-2}$}\\
\rule{0pt}{18pt} $T=T_{\text{min}}=\frac{\sqrt{(n-3)(n-1)}}{2\pi l}$ and $r_h=l\sqrt{\frac{n-3}{n-1}}$ & &\\
\rule{0pt}{18pt} At Hawking-Page phase transition: &\multirow{2}{*}{$R_{\text{HP}}=$} &\multirow{2}{*}{$-\frac{(n-1)(n-3)}{2}$~\cite{Wei2020}}   &\multirow{2}{*}{$-\frac{(n-3)2^n \pi^{n-2}}{\omega(n-2)^n}T_{\text{HP}}^{n-2}$}\\
\rule{0pt}{18pt} $T=T_{\text{HP}}=\frac{n-2}{2\pi l}$ and $r_h=l$ & &\\
\rule{0pt}{18pt} Ratio: &$\frac{R_{\text{HP}}}{R_{T_{\text{min}}}}=$     &$0$ &$\frac{n-3}{n-2}\left(\frac{n-3}{n-1}\right)^{n/2-1}$ and $\lim\limits_{n\rightarrow\infty}\frac{R_{\text{HP}}}{R_{T_{\text{min}}}}=\frac{1}{e}$\\
\hline
\end{tabular}
\caption{Comparison between our present results and those in Ref.~\cite{Wei2020}. Note that the results on the $R_N$ are only applicable to the coordinate space $\{T, V\}$, while the results on the $R^{(M)}$ are valid for the four different coordinate spaces mentioned in Section~\ref{sec2}.}
\label{tab}
\end{table}

Furthermore, a recent and intriguing development in understanding gravity has been the resurrection of the large dimension paradigm, in which  general relativity simplifies dramatically and its dynamics becoming trivial at all non-zero length scales away from the horizons of black holes~\cite{Emparan2013}. In the large dimension paradigm, a large body of works discuss various aspects of black holes, and the poignant observation, {\em any curvature is strongly localized near the horizon, and the spacetime quickly becomes
flat outside of it}, has been reported~\cite{Bhattacharyya2016,Emparan2015}. Ostensibly, one would go about this by computing all quantities as expansions of reciprocal dimension, then reading off the answer. In this work, we address the question of how the thermodynamic curvature of the AdS black hole behaves in this manner.

Due to the shrinking effect of the Euclidean time circle and of the area $\Omega_{n-2}$ of the spheres $S^{n-2}$ as $n\rightarrow\infty$, the Euclidean action of the black hole spacetime vanishes. In this case there is no trace of the Hawking-Page transition and the second-order phase transition at $n\rightarrow\infty$. One scheme is to rescale the free energy with an artificial fixed scale factor to realize these phase transition~\cite{Emparan2013}. The idea in this paper can be regarded as another scheme. We compute all quantities as expansions of dimension and read off the answer we need at $n\rightarrow\infty$. Then we can use the behavior of thermodynamic curvature to give a possible identification of the two phase transitions.

In the limit that the number of spacetime dimensions $n$ is infinite, for the Hawking-Page phase transition and the second-order phase transition, the behaviors of their corresponding temperature tend to be consistent, $\lim\limits_{n\rightarrow\infty} T_{\text{HP}}/T_{\text{min}}=1$, that is to say, there is no difference between the Hawking-Page phase transition and the second-order phase transition. While for the ratio of thermodynamic curvature along the curves of the Hawking-Page phase transition and the second-order phase transition, it becomes the natural constant $e$ at $n\rightarrow\infty$. This novel result reflects an essential difference between the Hawking-Page phase transition and the second-order phase transition of the AdS black hole. It also provides a universal and novel relationship between the Hawking-Page phase transition and the second-order phase transition of the AdS black hole in the large dimension paradigm.

Our current treatment can also be extended to charged black holes, where we need to pay attention to the criterion of the Hawking-Page phase transition when the black hole is charged. According to the idea of this paper, some universal relations of thermodynamic curvature of charged black hole in the various phase transitions need to be discussed in the future.

\section*{Acknowledgments}
Project funded by China Postdoctoral Science Foundation (Grant No. 2020M673460), National Natural Science Foundation of China (Grant Nos. 12047502 and 11947208), Scientific Research Program Funded by Shaanxi Provincial Science and Technology Department (2019JQ-081). This research is supported by The Double First-class University Construction Project of Northwest University. The authors would like to thank the anonymous referee for the helpful comments that improve this work greatly.

\end{document}